\documentstyle[12pt]{article}          

\newskip\humongous \humongous=0pt plus 1000pt minus 1000pt

\newif\ifdtup

\def\pr#1{#1^\prime}

\def\beq{\begin{equation}}
\def\eeq{\end{equation}}

\def\beqn{\begin{eqnarray}}
\def\eeqn{\end{eqnarray}}
\relax

\def\dotx{\dotx{\dot\overline{x}}}

\relax

\catcode`\@=11

\@addtoreset{equation}{section}
\def\theequation{\thesection\arabic{equation}}

\def\@normalsize{\@setsize\normalsize{15pt}\xiipt\@xiipt
\abovedisplayskip 14pt plus3pt minus3pt%
\belowdisplayskip \abovedisplayskip
\abovedisplayshortskip \z@ plus3pt%
\belowdisplayshortskip 7pt plus3.5pt minus0pt}

\def\small{\@setsize\small{13.6pt}\xipt\@xipt
\abovedisplayskip 13pt plus3pt minus3pt%
\belowdisplayskip \abovedisplayskip
\abovedisplayshortskip \z@ plus3pt%
\belowdisplayshortskip 7pt plus3.5pt minus0pt
\def\@listi{\parsep 4.5pt plus 2pt minus 1pt
     \itemsep \parsep
     \topsep 9pt plus 3pt minus 3pt}}

\@twosidetrue

\relax

\catcode`@=12

\catcode`\@=11

\def\section{\@startsection{section}{1}{\z@}{3.5ex plus 1ex minus
   .2ex}{2.3ex plus .2ex}{\large\bf}}

\def\thesection{\arabic{section}.}

\def\appendix{\setcounter{section}{0}
 \def\thesection{APPENDIX \Alph{section}:}
 \def\theequation{\Alph{section}.\arabic{equation}}}

\def\ps@headings{\def\@oddfoot{}\def\@evenfoot{}
\def\@oddhead{\hbox{}\hfill
 \makebox[.5\textwidth]{\raggedright\ignorespaces --\thepage{}--
 \hfill {}}}  
\def\@evenhead{\@oddhead}
\def\subsectionmark##1{\markboth{##1}{}}
}

\ps@headings

\catcode`\@=12

\def\figcap{\section*{Figure Captions\markboth
 {FIGURECAPTIONS}{FIGURECAPTIONS}}\list
 {Fig. \arabic{enumi}:\hfill}{\settowidth\labelwidth{Fig. 999:}
 \leftmargin\labelwidth
 \advance\leftmargin\labelsep\usecounter{enumi}}}
 \relax
\def\tablecap{\section*{Table Captions\markboth
 {TABLECAPTIONS}{TABLECAPTIONS}}\list
 {Table \arabic{enumi}:\hfill}{\settowidth\labelwidth{Table 999:}
 \leftmargin\labelwidth
 \advance\leftmargin\labelsep\usecounter{enumi}}}
 \relax
\def\reflist{\section*{References\markboth
 {REFLIST}{REFLIST}}\list
 {[\arabic{enumi}]\hfill}{\settowidth\labelwidth{[999]}
 \leftmargin\labelwidth
 \advance\leftmargin\labelsep\usecounter{enumi}}}
 \relax

\catcode`\@=11

\def\ps@headings{\def\@oddfoot{}\def\@evenfoot{}
\def\@oddhead{\hbox{}\hfill
 \makebox[.5\textwidth]{\raggedright\ignorespaces --\thepage{}--
 \hfill {}}}    
\def\@evenhead{\@oddhead}
\def\subsectionmark##1{\markboth{##1}{}}
}

\ps@headings

\relax

\relax
\def\pl#1#2#3{{\it Phys. Lett. }{\bf #1}(19#2)#3}
\def\zp#1#2#3{{\it Z. Phys. }{\bf #1}(19#2)#3}
\def\prl#1#2#3{{\it Phys. Rev. Lett. }{\bf #1}(19#2)#3}

\def\pr#1#2#3{{\it Phys. Rev. }{\bf #1}(19#2)#3}
\def\np#1#2#3{{\it Nucl. Phys. }{\bf #1}(19#2)#3}

\relax

\begin{document}
\def\theequation{\arabic{equation}}
\newcommand\sss{\scriptscriptstyle}
\newcommand\as{\alpha_{\sss S}}
\newcommand\ep{\epsilon}
\newcommand\aem{\alpha_{\rm em}}
\newcommand\refq[1]{$^{[#1]}$}
\newcommand\asb{\as^{(b)}}
\newcommand\epb{\overline{\epsilon}}
\newcommand\MSB{{\rm \overline{MS}}}
\renewcommand\topfraction{1}       
\renewcommand\bottomfraction{1}    
\renewcommand\textfraction{0}      
\setcounter{topnumber}{5}          
\setcounter{bottomnumber}{5}       
\setcounter{totalnumber}{5}        
\setcounter{dbltopnumber}{2}       
%
\begin{titlepage}
\nopagebreak
\vspace*{-1in}
{\leftskip 11cm
\normalsize
\noindent
\newline
CERN-TH.7023/93\\

}
\vfill
\begin{center}
{\large \bf On the $Q^2$ dependence of }

{\large \bf the measured polarized structure functions}
\vfill
{\bf G. Altarelli}, {\bf P. Nason\footnotemark}
\footnotetext{On leave of absence from INFN, Sezione di Milano, Milan, Italy.}
and
{\bf G. Ridolfi\footnotemark}
\footnotetext{On leave of absence from INFN, Sezione di Genova, Genoa, Italy.}
\vskip .3cm
{CERN TH-Division, CH-1211 Geneva 23, Switzerland}
\end{center}
\vfill
\nopagebreak
\begin{abstract}
{\small We analyse the available data on the polarized asymmetries $A_1$
for proton, neutron and deuteron targets. We use a homogeneous and updated
set of unpolarized structure functions to derive $g_1$ from $A_1$, and we
accurately correct for the scaling violations in order to obtain
$g_1(x,Q^2)$ with the same $Q^2$ for all $x$ values. The contribution to the
$Q^2$ evolution of a possible large gluon polarized density is also considered.
The implications for the Ellis-Jaffe and for the Bjorken sum rules are
discussed.}
\end{abstract}
\vfill
CERN-TH.7023/93
\newline
October 1993    \hfill
\end{titlepage}

New data on the polarized structure functions $g_1$ of
deuterium nuclei\refq{\ref{SMC}} and of
neutrons\refq{\ref{E142}} have recently been published,
which add to previous data on $g_1$ of protons\refq{\ref{SLAC},\ref{EMC}}. Of
special importance for the physical interpretation of the results is the
derivation from the data of the first moments related to the
Ellis-Jaffe\refq{\ref{EllisJaffe}} and the Bjorken\refq{\ref{Bjorken}} sum
rules. While the data on $g_1(x,Q^2)$ are collected at different values of
$Q^2$ for different values of $x$, the evaluation of moments in $x$ requires
the same value of $Q^2$ for all values of $x$. Thus, for a correct evaluation
of moments, in principle, one must apply corrections to the data points in
order to take the $Q^2$ evolution into account and reduce each data point in
$x$ to a common $Q^2$  value. As a first approximation to the solution of this
problem one can imagine to take advantage of the fact that, within the present
accuracy of the data, the primary measured quantity, the asymmetry
\beq
A_1=\frac
{\sigma^{\uparrow\downarrow}-\sigma^{\uparrow\uparrow}}
{\sigma^{\uparrow\downarrow}+\sigma^{\uparrow\uparrow}},
\eeq
shows no
appreciable $Q^2$ dependence\refq{\ref{SMC}-\ref{EMC}}.
Then, $g_1(x,Q^2)$ at fixed
$Q^2$ for all $x$ values is obtained from $A_1(x)$ through the relation
$A_1=g_1/F_1$, with $F_1$ being the unpolarized structure function
obtained from a fit of existing data. This procedure has been used in
refs.~[\ref{EllisKarliner},\ref{CloseRoberts}]. As confirmed by these
approximate analyses, the resulting corrections to the first moments are small
in comparison with the present experimental errors. However, in view of more
precise forthcoming data, it is interesting to collect the known results on the
$Q^2$ evolution of polarized parton
densities\refq{\ref{AltarelliParisi},\ref{Altarelli}} and describe a more
accurate method for reducing the set of data to the same $Q^2$ value for all
$x$ bins. In fact, an obvious shortcoming of the approximation of
refs.~[\ref{EllisKarliner},\ref{CloseRoberts}] is that the asymmetry $A_1$
is not
predicted to be a constant in $Q^2$ by the correct evolution equations, so
that, by assuming it to be a constant, one makes an error of the same order as
the effect under study. As the whole $Q^2$ correction is small with respect to
the present experimental errors, it is no surprise that the variation of $A_1$
with $Q^2$ is not discernible in the present data.

We describe in the following a more correct procedure that starts from
$A_1(x_i,Q^2_i)$, for the $i$-th experimental point,
constructs $g_1(x_i,Q^2_i)=A_1(x_i,Q^2_i) F_1(x_i,Q^2_i)$
and finally evolves $g_1(x_i,Q^2_i)$ into
$g_1(x_i,Q^2_0)$, with $Q^2_0$ being a suitable common value, for example the
average $Q^2$ value in the experiment. In passing, we explicitly compute the
predicted $Q^2$ dependence of the asymmetry $A_1$ for $p$, $n$ and $d$
targets, and check
that the available data are indeed compatible with it. The $Q^2$ evolution of
polarized quark densities also depends on the gluon polarized density. We
compute the evolution for two limiting cases. In the first case we assume a
negligible amount of polarized gluons. In the second case we start from a
polarized gluon density large enough to  entirely explain in terms of gluons
(with neglible polarized strange sea) the violation of the Ellis-Jaffe sum rule
observed by the EMC experiment on protons, according to the mechanism based on
the anomaly proposed in refs.~[\ref{AltarelliRoss},\ref{Efremov}]
and further discussed in
refs.~[\ref{CarlitzCollinsMuller}--\ref{AltarelliLampe}].

We consider the polarized structure functions
\beqn
\label{G1P}
&&g_1^p = \frac{1}{2}
\left[\frac{4}{9}\Delta u + \frac{1}{9}\Delta d + \frac{1}{9}\Delta s\right]
\\
\label{G1N}
&&g_1^n = \frac{1}{2}
\left[\frac{4}{9}\Delta d + \frac{1}{9}\Delta u + \frac{1}{9}\Delta s\right],
\eeqn
where
\beq
\Delta q = q_+ + \bar{q}_+ - q_- - \bar{q}_-
\eeq
and $q_+$ ($q_-$) is the density of the quark $q$ with
positive (negative) helicity in a proton with positive helicity.
We recall that, according to the mechanism discussed in
refs.~[\ref{AltarelliRoss}--\ref{AltarelliLampe}], the effective
polarized quark densities defined by eqs.~(\ref{G1P}) and (\ref{G1N})
are split into a part conserved by the two-loop $Q^2$ evolution equations,
which
is expected to be closely related to constituent quarks, and the gluon
component, which, due to the anomaly, does not decouple from the first moment
at large $Q^2$.

We have
\beqn
&&g_1^p =\phantom{-}g_1^{(3)} + g_1^{(8)} + g_1^{\sss (S)}
\\
&&g_1^n = -g_1^{(3)} + g_1^{(8)} + g_1^{\sss (S)},
\eeqn
where
\beqn
&&g_1^{(3)}    = \frac{1}{12}\left[\Delta u - \Delta d \right]
\\
&&g_1^{(8)}    = \frac{1}{36}\left[\Delta u + \Delta d - 2\Delta s\right]
\\
&&g_1^{\sss (S)} = \frac{1}{9} \left[\Delta u + \Delta d +  \Delta s\right].
\eeqn
The evolution equation for $\Delta q$ is given by \refq{\ref{AltarelliParisi}}
\beq
\label{evol}
Q^2\frac{d}{d Q^2} \Delta q =
\frac{\as(Q^2)}{2\pi}
\left[ \Delta q \otimes \Delta P_{qq}
+ 2\Delta g\otimes \Delta P_{qg} \right].
\eeq
Here $\Delta g=g_+ - g_-$, $g$ is the gluon density function, and
\beqn
&&\Delta  P_{qq}(z) = \frac{4}{3}\left(\frac{1+z^2}{1-z}\right)_+
\\
&&\Delta  P_{qg}(z) = \frac{z^2-(1-z)^2}{2}.
\eeqn
The convolution product $\otimes$ is defined as usual by
\beq
(f\otimes g)(x)=\int_x^1 \frac{dz}{z}f(z)g\left(\frac{x}{z}\right).
\eeq
Combining eqs.~(\ref{G1P}), (\ref{G1N}) and (\ref{evol}), we find
\beq
\label{evolG}
Q^2\frac{d}{d Q^2} g_1^{p,n} =
\frac{\as(Q^2)}{2\pi}
\left[ g_1^{p,n} \otimes \Delta P_{qq}
+ \frac{2}{3}\Delta g\otimes \Delta P_{qg} \right],
\eeq
or
\beq
\label{evolG2}
d g_1^{p,n} =
\frac{\as(Q^2)}{2\pi}d\log\frac{Q^2}{\Lambda^2}
\left[ g_1^{p,n} \otimes \Delta P_{qq}
+ \frac{2}{3}\Delta g\otimes \Delta P_{qg} \right].
\eeq
This equation can be approximately integrated by observing that
\beq
\frac{\as(Q^2)}{2\pi}d\log\frac{Q^2}{\Lambda^2}=
-\frac{1}{2\pi b} d\log\as(Q^2) +{\cal O}(\as^2),
\eeq
where $2\pi b=(33-2n_f)/6 = 9/2$. For a small displacement in $Q^2$, we
can neglect the $Q^2$ dependence of $g_1^{p,n}$ and $\Delta g$
on the r.h.s. of eq.~(\ref{evolG2}), and obtain
\beq
\label{evolG3}
g_1^{p,n}(x,Q^2)-g_1^{p,n}(x,Q_0^2) =
\frac{2}{9}\log\frac{\as(Q_0^2)}{\as(Q^2)}
\left[ g_1^{p,n} \otimes \Delta P_{qq}
+ \frac{2}{3}\Delta g\otimes \Delta P_{qg} \right]+{\cal O}(\as^2).
\eeq
Actually, in the numerical computations, we use the two-loop expression
of $\as(Q^2)$ in terms of $\Lambda$. We take
$\Lambda_{\overline{\sss MS}}(n_f=3)
=383^{+126}_{-116}$ MeV, which corresponds to $\as(m_{\sss Z}^2)=0.118\pm
0.007$.
With the definitions given above, taking into account the definition of
the $+$ distribution, one obtains
\beq
g_1^{p,n} \otimes \Delta P_{qq} =
\frac{4}{3}\int_x^1\frac{dz}{z}\frac{1+z^2}{1-z}
\left[\frac{1}{z}
g_1^{p,n}\left(\frac{x}{z}\right)-g_1^{p,n}(x)\right]
+\frac{4}{3}\left[x+\frac{x^2}{2}+2\log(1-x)\right]
\eeq
and
\beq
\Delta g\otimes \Delta P_{qg} =
\int_x^1 \frac{dz}{z} \frac{z^2-(1-z)^2}{2}\Delta g\left(\frac{x}{z}\right).
\eeq

An extreme parametrization of $\Delta g(x)$ has been proposed
in ref.~[\ref{AltarelliStirling}] for the EMC experiment:
\beq
\Delta g(x) = C x^{-0.3}(1-x)^7,
\eeq
with the normalization $C$ such that $\int_0^1 \Delta g(x) dx = 5$.
In the following, for each experiment we use the same form for
$\Delta g(x)$, but with a rescaled normalization given by
\beq
\int_0^1 \Delta g(x) dx = 5\frac{\as(Q^2_{\sss EMC})}{\as(Q^2_{\sss EXP})},
\eeq
where $Q^2_{\sss EMC}=10.7$ GeV$^2$ and $Q^2_{\sss EXP}$
are the average values of $Q^2$ for the EMC and for the experiment under
consideration, respectively.
This amount of polarized gluons is extreme in the sense that all the
EMC deviation from the Ellis-Jaffe sum rule is attributed to the effect of
gluons, while some amount of polarized strange sea is indeed plausible. Also,
the above polarized gluon distribution nearly saturates the positivity
bound from the known unpolarized gluon density especially at large $x$
[\ref{AltarelliStirling}].

In the original treatment of the data\refq{\ref{SMC}-\ref{EMC}}, in order to
obtain the values of $g_1$, the measured asymmetry $A_1(x)$ is multiplied by
$F_1(x,Q^2_0)$, where $Q^2_0$ is the average $Q^2$ of the experiment. In the
present context, obviously we cannot tolerate this neglect of an important
contribution to the $Q^2$ dependence. As already stated in the introduction, we
construct $g_1(x_i,Q^2_i)=A_1(x_i,Q^2_i) F_1(x_i,Q^2_i)$. At this stage we can
implement, for all experiments, an updated form for the unpolarized structure
functions that takes also the recent measurements by the NMC\refq{\ref{NMCf2}}
into account. For this purpose we use the
fit to the structure function $F_2$ from ref.~[\ref{NMCf2}] and the fit to
$R=F_2/(2xF_1)-1$ from ref.~[\ref{RSLAC}]. Starting from
the resulting values of $g_1(x_i,Q^2_i)$, we evolve each point up to
$g_1(x_i,Q^2_0)$. For the evolution we use a fit to the data points
$g_1(x_i,Q^2_i)$ as an input for $g_1(x)$ in the r.h.s. of eq.~(\ref{evolG3})
and consider both the cases with and without gluons. The modified experimental
points for $g_1$,
and the corresponding ones corrected for the evolution to $Q^2_0$,
chosen as the average $Q^2$ of the experiment, are shown in
fig.~\ref{g1data}. In the same
figure we also show a fit to the data, (solid line), corrected for the
evolution with $\Delta g =0$. Our fits are given by
\beqn
g_1^{p}(x,Q^2_0)&=& x^{0.2} (1-x)^3 (0.589+1.07x-1.29x^2-0.792x^3)
\\
g_1^{d}(x,Q^2_0)&=& x^{0.2} (1-x)^6 (-0.670+15.0x-46.0x^2+43.4x^3)
\\
g_1^{n}(x,Q^2_0)&=& x^{0.2} (1-x)^6 (-0.358-1.10x+12.7x^2-16.5x^3).
\eeqn
By $d$ we actually mean $(p+n)/2$, as the corresponding nuclear correction
factor has been taken into account\refq{\ref{SMC}}.
The extrapolation at low $x$ outside the measured
range is also shown, based on fitting only the last two measured points by $B
x^{0.2}$ (i.e. the same $\alpha$ as in the previous fit). Similarly the
extrapolation at large $x$, based on $D (1-x)^3$ fitted on only the last
measured point, is also displayed.

In fig.~\ref{a1emc}
we compare the computed $Q^2$ dependence of the measured asymmetry
for protons with the EMC data\refq{\ref{EMC}}. For each bin in $x$ the two
curves refer to the computed $Q^2$ dependence at the lowest and highest point
in $x$. The $Q^2$ dependence of $A_1(x,Q^2)$ is evaluated by using the
evolution
equations for $g_1(x,Q^2)$ according to the above formalism and the already
mentioned ($Q^2$-dependent) fit to the unpolarized structure functions in order
to obtain $F_1(x,Q^2)$. The results with and without gluons are
plotted separately.
We see that the difference associated to the gluons in the evolution
is small. This is because the evolution equations at $x_0$ are only sensitive
to the parton densities at $x>x_0$: the gluon density, although having a large
first moment, being concentrated at small
$x$, has little influence on the evolution in the measured range of $x$. We
confirm that within the present accuracy of the data no evidence for
the $Q^2$ dependence can be seen. But we also see that the effects of the $Q^2$
dependence cannot be ignored in a more precise analysis. In fig.~\ref{a1dp} the
predicted $Q^2$ dependence of the neutron and deuterium asymmetries is shown.

The first moments of $g_1$ for $p$, $n$ and $d$ were obtained by three
different methods:
\begin{itemize}
\item (a) We sum the experimental points bin by bin, excluding the first
and the last bins. We then add to this value the integral of the small and
large
$x$ extrapolations over the remaining $x$ region. This way of treating the
end points is safer, since the functions $g_1$ are rapidly varying there,
especially near $x=1$.
\item (b) We integrate our fits of the data over the experimental
range excluding the first and the last bins. The remaining region is treated
as in (a).
\item (c) We simply integrate our fits over the full $x$ range.
\end{itemize}
The results corresponding to the different integration methods,
with and without the corrections for the $Q^2$ scaling violations, and with and
without gluons in the evolution equations, are collected in table~1 for $p$,
$n$, $d$ and $p-n$, the last being the combination of relevance for testing the
Bjorken sum rule. For protons, the
EMC\refq{\ref{EMC}} and SLAC\refq{\ref{SLAC}} data have been combined.
It can be seen that the differences between moments computed by
different methods are well within the errors quoted by the collaborations.
We prefer procedure (c) of
integrating the overall fit over the complete range. In fact the sum bin by bin
does not take the differences in the errors associated with the different bins
into proper account, and the extrapolation based on only the last points
overemphasizes their significance. Nevertheless, we consider it important to
keep in mind the size of the changes from one method to the other.

\begin{table}
\begin{center}
\begin{tabular}{|l||c|c|c||c|c|c||c|c|c|} \hline
& \multicolumn{3}{c||}{I}
& \multicolumn{3}{c||}{II}
& \multicolumn{3}{c|}{III}
\\ \hline
& (a)
& (b)
& (c)
& (a)
& (b)
& (c)
& (a)
& (b)
& (c)
\\ \hline \hline
$I(p)$& 0.125 & 0.122 &  0.119 & 0.132 & 0.126 & 0.119 & 0.131 & 0.127 & 0.119
 \\ \hline
$I(d)$& 0.029 & 0.029 &  0.031 & 0.034 & 0.034 & 0.035 & 0.031 & 0.031 & 0.034
 \\ \hline
$I(n)$&-0.023 &-0.023 & -0.022 &-0.023 &-0.023 &-0.023 &-0.022 &-0.022 &-0.020
 \\ \hline
\hline
& \multicolumn{9}{c|}{Bjorken sum rule}
\\ \hline \hline
EMC/SMC  & 0.192 & 0.185 & 0.175 & 0.196 & 0.185 & 0.167 & 0.200 & 0.191 &
0.171
 \\ \hline
EMC/E142 & 0.148 & 0.144 & 0.141 & 0.155 & 0.149 & 0.141 & 0.153 & 0.148 &
0.140
 \\ \hline
SMC/E142 & 0.104 & 0.104 & 0.106 & 0.114 & 0.114 & 0.115 & 0.105 & 0.105 &
0.108
 \\ \hline
\end{tabular}
\caption[]{\label{uno}
First moments of $g_1$ for $p$, $d$ and $n$. The group of columns labelled
I correspond to uncorrected data, those labelled II (III) to data corrected for
evolution with $\Delta g=0$ ($\Delta g$ as in eq.~(19)). The columns (a),
(b) and (c) refer to the different integration methods described in the text.
}
\end{center}
\end{table}

The results in table~1 are obtained with the central value $\Lambda=383$ MeV.
The effect of varying $\Lambda$ in the range given above, $267$ to $509$ MeV,
amounts to a relative error of the order of 20 to 40\% of the computed
correction due to $Q^2$ evolution. However, because of the smallness
of the evolution correction, the absolute effect on the first
moments $I(p)$, $I(d)$ and $I(n)$ is always less than $\pm 0.0012$.
According to the above discussion, for the first moments $I(p)$, $I(d)$
and $I(n)$ we take the results from columns (c) of table~1.
The associated errors are taken to be
identical to those quoted in the experimental
papers, as is appropriate, given our indicative purposes. We obtain
the following results:
\beqn
\label{moments}
I(p) &=& [0.119,0.119,0.119] \pm 0.018   \;\;\; (0.126)
\\
I(d) &=& [0.031,0.035,0.034] \pm 0.027   \;\;\; (0.0245)
\\
I(n) &=& [-0.022,-0.023,-0.020] \pm 0.011 \;\;\; (-0.022),
\eeqn
where in square brackets we give in a sequence the raw data, the results
from the evolution
with only quarks, and finally those with quarks and gluons. The values in round
brackets are the raw data values quoted by the experimental collaborations
(with the error which is also shown).
We see that the effect
of the $Q^2$ evolution is small with respect to the ambiguities associated
with both the fitting procedure and the experimental errors.

For the Bjorken difference $I(p)-I(n)$, obtained from the three possible
combinations of experiments, we obtain, in the same notation as above:
\beqn
\label{sumrule}
\label{bj1}
{\rm EMC/SMC}  &:& [0.175,0.167,0.171] \pm 0.060   \;\;\; (0.200)
\\
\label{bj2}
{\rm EMC/E142} &:& [0.141,0.141,0.140] \pm 0.022   \;\;\; (0.148)
\\
\label{bj3}
{\rm SMC/E142} &:& [0.106,0.115,0.108] \pm 0.058   \;\;\; (0.093).
\eeqn
Note that the result on the Bjorken sum rule depends on the amount of polarized
gluons. One should not be surprised by this effect. It arises because the
gluons contribute to correct the individual data points and the average $Q^2$
is different for each experiment. Also note that the first moments are
independent of $Q^2$ in leading order: once the integral has been computed for
a given experiment at $Q^2_{\sss EXP}$, within the present accuracy,
it can be combined with the results of other experiments at different $Q^2$
values. As a consequence, for each entry in eqs.~(\ref{bj1})-(\ref{bj3}) we
cannot specify
the relevant $Q^2$ within the range defined by the two experiments. In fact,
the predicted difference in value for the Bjorken integral is of order
$(\as/\pi)^2$, which is beyond the present accuracy.

In fig.~\ref{bjth} we compare the experimental numbers for the Bjorken sum
rules, given in eqs.~(\ref{bj1})-(\ref{bj3}), with the $Q^2$-dependent
theoretical prediction, including corrections up to order $\as/\pi$ or
$(\as/\pi)^3$ calculated in ref.~[\ref{Vermaseren}]. The higher-twist
correction, as computed in ref.~[\ref{Ji}], is negligible in the $Q^2$ range
shown in fig.~\ref{bjth}.

In conclusion, we have studied the effect of scaling violation on the
determination of the polarized structure function $g_1$ for $p$, $n$ and $d$
targets, and we have proposed a procedure to correct for this effect. Our
results for the Bjorken sum rule are summarized in fig.~\ref{bjth}. We can see
that our results do not differ much from the results quoted by the SMC and E142
collaborations. One important aspect of fig.~\ref{bjth} are the large error
bars on the $Q^2$ value associated with each determination. We have taken these
error bars to be the range of the average $Q^2$ of each pair of experiments.
When this uncertainty range is taken into proper account, we see that the
experimental results are consistent with the theoretical prediction, although
on the low side. Unfortunately, we also see that the result obtained using the
E142 data over lap with a region of moderately small $Q^2$, where the
theoretical prediction in strongly unstable. The combination of EMC and SMC
data is in a better position from this point of view, but in this case the
errors are much larger.

\noindent {\bf Acknowledgements}

We thank V.~Breton, E.W.~Hughes, V.W.~Hughes, R.~Piegaia and M.~Werlen for
kindly providing us with useful information on the experimental data.

\newpage
\begin{reflist}
\item\label{SMC}
   B.~Adeva et al., SMC, \pl{B302}{93}{533}.
\item\label{E142}
   P.L.~Anthony et al., E142 Collaboration, \prl{71}{93}{959}.
\item\label{SLAC}
   M.J.~Alguard et al., E80 Collaboration, \prl{37}{76}{1261},
   {\bf 41}(1978)70;\\
   G.~Baum et al., E130 Collaboration, \prl{51}{83}{1135}.
\item\label{EMC}
   J.~Ashman et al., EMC, \pl{B206}{88}{364}; \np{B328}{89}{1}.
\item\label{EllisJaffe}
   J.~Ellis and R.L.~Jaffe, \pr{D9}{74}{1444}; {\bf D10}(1974)1669.
\item\label{Bjorken}
   J.D.~Bjorken, \pr{148}{66}{1467}; {\bf D1}(1970)1376.
\item\label{EllisKarliner}
   J.~Ellis and M.~Karliner, \pl{B313}{93}{131}.
\item\label{CloseRoberts}
   F.E.~Close, Rutherford preprint RAL-93-034 (1993);\\
   F.E.~Close and R.G.~Roberts, \pl{B316}{93}{165}.
\item\label{AltarelliParisi}
   G.~Altarelli and G.~Parisi, \np{B126}{77}{298}.
\item\label{Altarelli}
   G.~Altarelli, {\it Phys. Reports} {\bf 81}(1982)1.
\item\label{AltarelliRoss}
   G.~Altarelli and G.G.~Ross, \pl{B212}{88}{391}.
\item\label{Efremov}
   A.V.~Efremov and O.V.~Teryaev, Dubna preprint E2-88-287,
   {\it Czech. Hadron Symp.} {\bf 302} (1988).
\item\label{CarlitzCollinsMuller}
   R.D.~Carlitz, J.C.~Collins and A.H.~Mueller, \pl{B214}{88}{229}.
\item\label{AltarelliStirling}
   G.~Altarelli and W.J.~Stirling, {\it Particle World} {\bf 1}(1989)40.
\item\label{AltarelliLampe}
   G.~Altarelli and B.~Lampe, \zp{C47}{90}315.
\item\label{NMCf2}
   P.~Amaudrus et al., NMC, \pl{B295}{92}{159}.
\item\label{RSLAC}
   L.W.~Whitlow et al., \pl{B250}{90}{193}.
\item\label{Vermaseren}
   S.A.~Larin and J.A.M.~Vermaseren, \pl{B259}{91}{345}.
\item\label{Ji}
   X.~Ji and P.~Unrau, preprint MIT-CTP-2232 (1993).
\end{reflist}

\newpage
\begin{figcap}
\item\label{g1data}
Effect of the $Q^2$ evolution correction on the data for the structure function
$g_1(x,Q^2)$. Both the raw data and the data evolved to the average $Q^2$ of
each experiment (with and without the gluon contribution) are shown. The solid
curve is our fit of the data evolved without gluon. The dashed curves represent
the extrapolation of the data outside the measured range, as described in
the text.
\item\label{a1emc}
Computed $Q^2$ dependence of the proton asymmetry compared with the EMC
(squares) and E130-E80 (crosses) data.
The low (high) curves correspond to the lower (upper) edge of the $x$ bin.
\item\label{a1dp}
Computed $Q^2$ dependence of the deuteron and neutron asymmetry,
with (a) and without (b) the gluon contribution.
\item\label{bjth}
Theoretical prediction for the Bjorken sum rule as a function of $Q^2$.
The dotted curves are obtained using the ${\cal O}(\as)$ formula, for
$\Lambda=267$, $383$ and $509$ MeV. The solid lines include corrections
up to ${\cal O}(\as^3)$, for the same values of $\Lambda$.
The experimental values are also shown.
\end{figcap}
\end{document}